\documentclass[structabstract]{aa}  
\usepackage{graphicx}
\usepackage{txfonts}
\begin{document}

   \title{The intracluster magnetic field power spectrum in Abell 665}

   \subtitle{}
\author{  V. Vacca\inst{1}
          \and
          M. Murgia\inst{2,3}
          \and
          F. Govoni\inst{2}
          \and
          L. Feretti\inst{3}
         \and
          G. Giovannini\inst{3,4}
         \and
          E. Orr\`u\inst{5}
         \and 
          A. Bonafede\inst{3,4}   
          }
\institute{
              Dipartimento di Fisica,
           Universit\`a degli studi di Cagliari, Cittadella Universitaria, I--09042 Monserrato (CA), Italy
           \and
              INAF - Osservatorio Astronomico di Cagliari,
              Poggio dei Pini, Strada 54, I--09012 Capoterra (CA), Italy 
           \and   
              INAF - Istituto di Radioastronomia, 
              Via Gobetti 101, I--40129 Bologna, Italy
           \and 
              Dipartimento di Astronomia, 
              Univ. Bologna, Via Ranzani 1, I--40127 Bologna, Italy
          \and
              Institute for Astro$-$ and Particle Physics, University of Innsbruck, Technikerstr. 25, 6020 Innsbruck, Austria
                }

   \date{Received MM DD, YY; accepted MM DD, YY}

 %
  \abstract
  {}
   {The goal of this work is to investigate the power spectrum of the magnetic field associated with the giant radio halo 
in the galaxy cluster A665.}
   {For this, we present new deep Very Large Array total intensity and polarization observations at 1.4\, GHz. We simulated Gaussian random three-dimensional turbulent magnetic field models to reproduce the observed radio halo emission. By comparing observed and synthetic radio halo images we constrained the strength and structure of the 
intracluster magnetic field. We assumed that the magnetic field power spectrum is a power law with a Kolmogorov index and we imposed a local 
equipartition of energy density between relativistic particles and field.}
{Under these assumptions, we find that
the radio halo emission in A665 is consistent with a central magnetic field strength of about 1.3 $\mu$G. To explain the azimuthally averaged radio brightness profile, the magnetic field energy density should decrease following the thermal gas density, leading to an averaged magnetic field strength over the central 1 Mpc$^3$ 
of about 0.75 $\mu$G. From the observed brightness fluctuations of the radio halo, we infer that the outer scale
 of the magnetic field power spectrum is $\sim$ 450 kpc, and the corresponding magnetic field auto-correlation 
length is $\sim$100 kpc.}
  {}
 

   \maketitle

\begin{table*}
\caption{Details of the VLA observations of Abell 665.}             
\label{observation_details}      
\centering          
\begin{tabular}{c c c c c c c c}     
\hline\hline       
Obs. pointing, RA     &Obs. pointing, DEC    & Obs. frequency, $\nu$ & Bandwidth & Config. & Date & Time & Project  \\
(J2000) & (J2000) & (MHz) &  (MHz)  & ~ & ~ & (h) & ~ \\
\hline
08:30:53.000  &     +65:51:30.000  &   1465, 1415  &    25 & C & 24-Jul-2005 & 13 & AG690 \\
08:30:54.830  &     +65:49:54.800  &   1365, 1435  &    50 & D & 14-Jul-1996 & 5 & AF304 \\
\hline    
\end{tabular}
\end{table*}

\section{Introduction}

In the hierarchical model of structure formation, clusters of galaxies
undergo several merger events during their lifetimes. Shocks and the
turbulence associated with a major cluster merger event are thought to
accelerate particles and compress magnetic field in the intracluster
medium (e.g. Roettiger et al. 1999). 

The existence of relativistic particles and a
magnetic field is expected to lead to large-scale diffuse
emission, which is not associated with
individual host galaxies but is a general property of the intracluster medium. 
Radio observations reveal the presence of sources known as halos. 
Radio halos are faint ($\sim$ 1$\mu$Jy arcsec$^{-2}$ at 1.4 GHz),
steep-spectrum\footnote{$S(\nu)\propto \nu ^{-\alpha}$} ($\alpha\gtrsim$1) synchrotron sources located at the center of galaxy clusters 
(e.g. Feretti \& Giovannini 2008, Ferrari et al. 2008). 
As recently shown by Cassano et al. (2007) and Murgia et al. (2009), these diffuse radio sources can have quite different 
length scales, but the largest halos are the most luminous, in such a way that all these sources may have  
similar synchrotron emissivity.

About 30 radio halos nowadays are known, and all of them are found to be strictly 
related to intense merger activity (e.g. Schuecker et al. 2001; Buote 2001; Govoni et al. 2001).
Indeed, the most luminous radio halos are mainly associated to hot and 
massive clusters. Cluster merger events are expected to release a significant amount of energy
in the intracluster medium. This energy is injected on large spatial scales, and then turbulent
cascades may be generated (e.g. Brunetti \& Lazarian 2007).

In the past years, progress has been made by analyzing
 the state of turbulence of the intracluster medium. Signature of turbulence
have been found by analyzing the gas pressure of the Coma cluster,
where the pressure fluctuations are consistent with a Kolmogorov power spectrum (Schuecker et al. 2004).
Turbulence in the intracluster medium has also been studied in the radio band through Faraday rotation
measures. 
Recent results show that the magnetic field power spectrum
can be estimated if very detailed rotation-measure images
of cluster radio galaxies are available
(En{\ss}lin \& Vogt 2003; Murgia et al. 2004;  
Vogt \& En{\ss}lin 2005; Govoni et al. 2006; Guidetti et al. 2008; 
Laing et al. 2008). Typically, rotation measure images of cluster radio galaxies
permit investigating the fluctuations of the intracluster magnetic field
below a spatial scale of about $50\div 100$ kpc. On the other hand, radio halo
images reveal that the intracluster magnetic field is spread over Mpc scales.
Thus, it would be important to investigate the magnetic field turbulence 
over such a large volume of space.
In fact, although radio halos are typically unpolarized, in  
A2255 (Govoni et al. 2005) and MACS J0717.5+3745 (Bonafede et al. 2009),
 a polarized signal associated to the radio halo has also been detected. These
observations indicate that, at least in these clusters, the intracluster magnetic 
fields can be ordered on scales of a hundred kpc. 

In this paper we present a study of the
magnetic field power spectrum in the galaxy cluster A665,
which contains a known radio halo.
Following Tribble (1991) and Murgia et al. (2004), 
we propose to study the cluster magnetic field strength and structure on the basis
of the radio halo properties.
Indeed, information on the cluster magnetic field 
can be derived from detailed images of the radio halo, since
the halo brightness fluctuations and the polarization level 
are strictly related to the intracluster magnetic-field power spectrum.
For example, lack of polarization and a smooth and regular surface brightness may indicate 
that the cluster magnetic field is tangled on smaller scales than the resolution
of the radio images, while a disturbed radio morphology and the presence of polarization 
could be related to a magnetic field ordered on larger scales than the observing beam (Govoni et al. 2005).

We investigate the total intensity and the polarization properties 
of the A665 radio halo by means of a new deep VLA observation 
at 1.4 GHz in C configuration and previously published data at 1.4 GHz in D array 
(Giovannini \& Feretti 2000).
The power spectrum of the intracluster magnetic field fluctuations
is constrained by comparing the data with the expectations of
simulated random 3-dimensional magnetic field models characterized by different 
power spectra.

The structure of this paper is as
follows. In Sect. 2 we give a brief description of the cluster of galaxies A665. In Sect. 3 we
present the radio observations and describe the data reduction. In Sect. 4 we develop the
magnetic field modeling. In Sect. 5 we report the simulation results and the
radio halo analysis. In Sect. 6 we analyze the depolarization of the discrete radio sources. Finally, 
in Sect. 7 we summarize our conclusions.

Throughout this paper we adopted a $\Lambda$CDM cosmology with $H_0=71$ km s$^{-1}$ Mpc$^{-1}$,
$\Omega_{m}=0.27$, and $\Omega_{\Lambda}=0.73$. At the distance of A665 ( z=0.1829,  G{\'o}mez et al. 2000), 
1\arcsec\, corresponds to 3.04\,kpc.

\section{The cluster of galaxies Abell 665}
\label{Abell665}

 \begin{figure*}[ht]
   \centering
  \includegraphics[width=13cm]{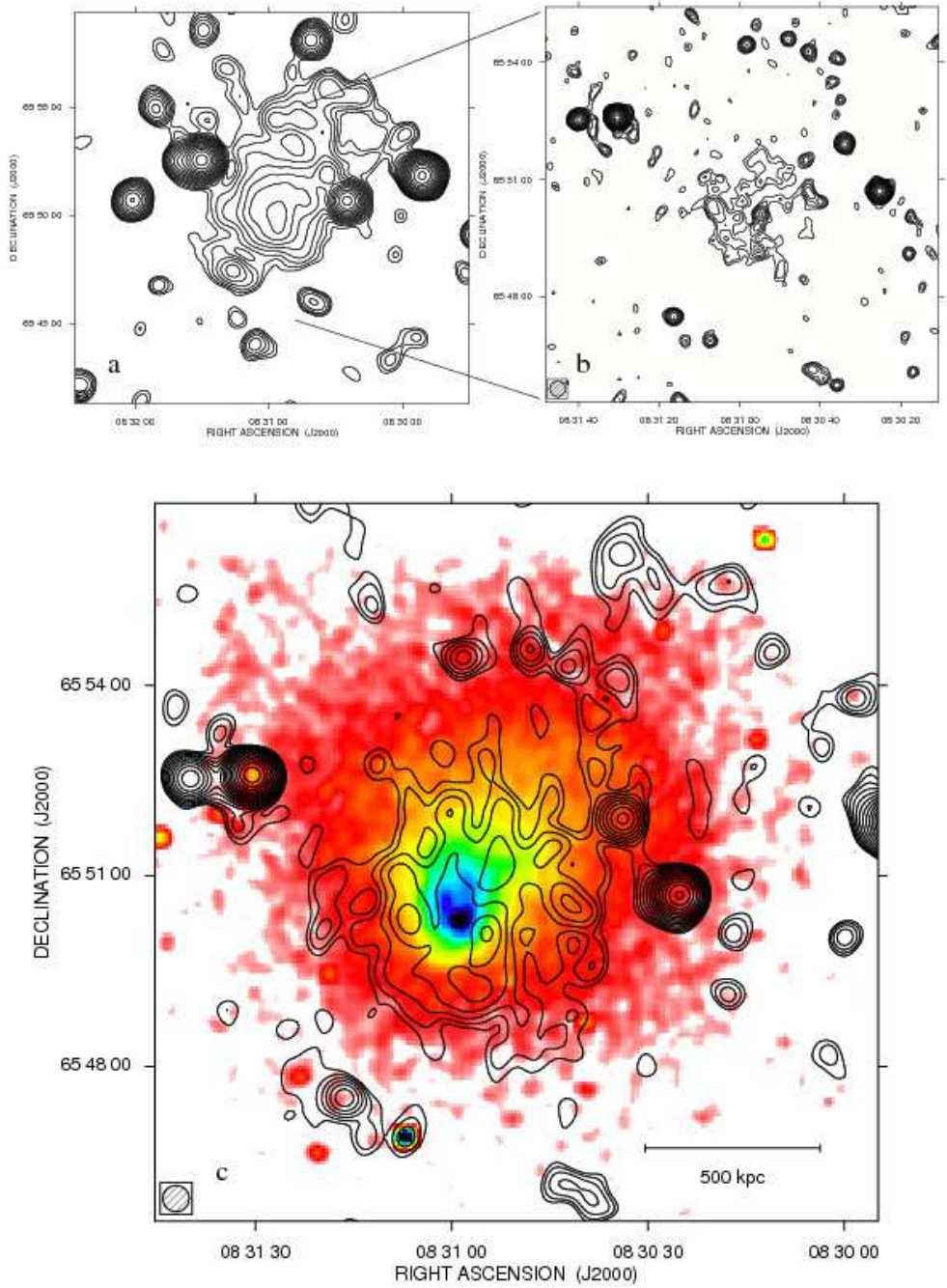}
     \caption{(a): Total intensity radio contours of A665 at 1.4 GHz (VLA in D configuration) with an
FWHM of $55\arcsec \times 55\arcsec$. The first contour level is drawn at 180 $\mu$Jy/beam,
and the rest are spaced by a factor $\sqrt{2}$. The sensitivity (1$\sigma$) is 60 $\mu$Jy/beam.
(b): Total intensity radio contours of A665 at 1.4 GHz (VLA in C configuration) with a
FWHM of $15\arcsec \times 15\arcsec$.
The first contour level is drawn at 75 $\mu$Jy/beam,
and the rest are spaced by a factor $\sqrt{2}$.
The sensitivity (1$\sigma$) is 25 $\mu$Jy/beam.
(c): Total intensity radio contours of A665 at 1.4 GHz (combining VLA data in C and D configuration) with an
FWHM of $25\arcsec \times 25\arcsec$. The first contour level is drawn at 135 $\mu$Jy/beam,
and the rest are spaced by a factor $\sqrt{2}$. The sensitivity (1$\sigma$) is 45 $\mu$Jy/beam. The contours
 of the radio intensity are overlaid on the \emph{Chandra} X-ray image. The adaptively smoothed X-ray image is in the 0.8-4 keV band 
(Govoni et al. 2004).}
              \label{A665}
    \end{figure*}

The cluster of galaxies A665 is well known to contain an extended diffuse radio halo first observed
by Moffet \& Birkinshaw (1989) and successively confirmed by Jones \& Saunders (1996). At an angular resolution
of about 1\arcmin, the radio halo is asymmetric because it is elongated in the southeast-northwest direction (Giovannini \& 
Feretti 2000).

The total halo flux density at 1.4\,GHz is $43.1\pm0.8$\,mJy, while at 0.327\,GHz the flux
density is $197\pm6$\,mJy. This results in an averaged spectral index of
$\alpha_{0.3}^{1.4}=1.04\pm 0.02$ (Feretti et al. 2004). Like all clusters containing a radio halo, A665 shows signatures of a major merger. 
Optical (e.g. G{\'o}mez et al. 2000) and X-ray (e.g. Hughes 
\& Birkinshaw 1994) studies have revealed a disturbed dynamical state.

\emph{Chandra} satellite data have provided a detailed gas temperature map
revealing a shock region with a temperature jump  from 8 keV to 15 keV, located near the southern boundary of the 
radio halo (Markevitch \& Vikhlinin 2001; Govoni et al. 2004). The radio morphology seems to follow the X-ray 
elongation, trailing the shock wave direction.
A spectral analysis of \emph{Chandra} data by  Million \& Allen (2009) indicates the possibility of 
nonthermal (or quasi-thermal) X-ray emission, likely due to inverse Compton scatter between relativistic
electrons responsible for radio halo emission and cosmic microwave background photons.

\section{Radio observations and data reduction}
\label{Radio observations and data reduction}
We present a new deep, polarization sensitive, full-synthesis observation of A665 at 1.4\,GHz obtained
using the Very Large Array (VLA) in C configuration. The details of the observation are
summarized in Table\,\ref{observation_details}. The data reduced following standard 
procedures using the NRAO's Astronomical Image Processing System (AIPS) package.

The nearby phase calibrator 0834+555 was observed 
over a wide range in parallactic angle 
to separate the source polarization properties 
from the feed parameters. The radio source 3C286 was used both as 
primary flux density calibrator and as reference for the absolute polarization angles.  
Radio interferences were carefully excised, and several cycles of self-calibration and CLEAN were applied 
to remove residual phase variations.
A circular beam of 15\arcsec\, was used to restore  
the final images of total intensity $I$ and Stokes parameters 
$Q$ and $U$. Images of polarized intensity $P=\sqrt{Q^2+U^2}$ (corrected for the positive bias) 
and fractional polarization $FPOL=P/I$ were derived from the $I$, $Q$, and $U$ images.
The $I$ image has an rms noise level of $\sigma_I\simeq 25$ $\mu$Jy/beam, while
the $P$ image has an rms noise level of $\sigma_P\simeq 15$ $\mu$Jy/beam.  

In order to improve $uv$-coverage and sensitivity to the diffuse emission 
but still keeping a good angular resolution, we combined the new C configuration data with 
the D configuration data by Giovannini \& Feretti (2000), see Table\,\ref{observation_details}. The D configuration data were been calibrated in phase and amplitude 
by using the sources 0841+708 and 3C48, respectively.
A circular beam of 55\arcsec\, was applied to restore  
the final total intensity image $I$, which has a noise level of 60 $\mu$Jy/beam. The combined C+D data were then self-calibrated 
to produce a final total intensity image. This image was restored with a circular beam of 25\arcsec\, and has 
a noise level of 45 $\mu$Jy/beam. Since the D configuration data were
 not polarization sensitive, no $Q$ and $U$ images were produced with the combined
 C+D data set. The total intensity images at different resolution of the radio emission in A665 are shown in Fig.\,\ref{A665}.

The central regions of the galaxy cluster are permeated by a low-surface brightness
radio halo which is surrounded by many discrete radio sources. The full extent of the radio halo is readily visible in the D configuration image shown
in Fig.\,\ref{A665}a. However, the comparatively low resolution of this image
does not permit easily distinguishing the contribution of the discrete radio sources from the diffuse radio halo.

The new C configuration image at a resolution of 15\arcsec\,is shown in Fig.\,\ref{A665}b. 
Here we can clearly separate the halo emission from the discrete radio sources. In particular, 
there are a few embedded sources in the central regions, and at least part of the
northwestern elongation of the halo seen at lower resolution comes from a blending of several 
discrete sources distributed, in projection, along an arc-like structure.
At the sensitivity level of our C configuration observation, we did not detect polarized emission
from the radio halo. Since the innermost regions of the halo have an average brightness of $I\simeq 100$ $\mu$Jy/beam, 
we can derive a 1$\sigma$ upper limit to the fractional polarization of $FPOL\le\sqrt{\sigma_{P}^2/(I^2-\sigma_{I}^2)}\simeq 15\%$. The polarization of the discrete sources in the field of view is discussed in Sect.\ref{Radio galaxy emission in A665}.

Finally, In Fig.\ref{A665}c we present the C+D configuration image at a resolution of 25\arcsec (corresponding to 75 kpc at the cluster distance). The radio iso-contours\footnote{ The reader can refer to Fig.\,\ref{simulations}a for an alternative visualization with contours and color scale of the radio emission at 1.4\,GHz.} are overlaid on the adaptively smoothed \emph{Chandra} X-ray image in 
the 0.8-4 keV band (Govoni et al. 2004). The X-ray and radio peaks do not overlap precisely. 
The distance between them is about 45\arcsec. The radio halo emission peaks at  RA=$08^{h}31^{m}7^{s}$ and DEC=$65^{d}50^{m}21^{s}$, while the X-ray peak coordinates are  
RA=$08^{h}30^{m}59^{s}$ and DEC=$65^{d}50^{m}21^{s}$.

\section{Magnetic field modeling}
\label{Characterization of the diffuse radio emission in A665}
The halo brightness fluctuations and the polarization level can be 
related to the intracluster magnetic-field power spectrum.  
Following the approaches proposed by Tribble (1991) and Murgia et al. (2004), 
we analyzed the radio images presented in the previous section to constrain
 the intracluster magnetic field in A665.
We proceeded in two steps. First, we performed a 2-dimensional
analysis to determine the radial scaling of the magnetic field, i.e. the average
magnetic field strength at the cluster center and its radial
decline. Second, we focused on the radio-halo total intensity 
and polarization fluctuations with respect to the average radial
profiles to determine their power spectrum with the aid of 3-dimensional 
numerical simulations.

\subsection{The radial scaling of the equipartition magnetic field strength}
\label{2D}
In the following 2-dimensional analysis, we want to infer the average magnetic field strength 
at the cluster center and possibly its scaling with the thermal gas density (see also Murgia et al. 2009). 
The observed radio halo intensity is directly related to the integral of
the magnetic field and relativistic particle distribution along the line-of-sight. 
However, since the synchrotron emissivity traces the product of electron and 
magnetic field energy densities, disentangling the two contributions is not 
possible from radio observations alone.
In this work we assume that the electron energy spectrum is a power law of the
form

\begin{equation}
N(\epsilon, \theta)=N_0\epsilon ^{-\delta} \frac{\sin\theta}{2}
\label{CRe}
\end{equation}
with the energy $\epsilon$ ranging from $\gamma_{min}m_ec^2$ and $\gamma_{max}m_ec^2$, with $\gamma_{max}\gg\gamma_{min}$.  Such an energy spectrum  is expected in the steady state of a continuous injection of power-law distributed cosmic-ray electrons subject to radiative losses. The 
index $\delta$ is related to the observed spectral index throughout the relation $\delta=2\alpha+1$. The relativistic
electrons are supposed to have an isotropic distribution of the pitch angle, $\theta$, between the direction of their
velocity and the magnetic field. We set $\delta=3$, in agreement with the observed spectral index $\alpha_{0.3}^{1.4}\simeq 1$
 (see Sect.\,\ref{Abell665}).
We set the normalization $N_0$ supposing a perfect equipartition condition between magnetic field and relativistic particle 
energy density at each point in the intracluster medium. In this case the radio source is on a minimum energy
condition\footnote{ The energy densities of the magnetic field, $u_B$, and cosmic ray particles, $u_{cr}$, are exactly  equal if the minimum energy condition is calculated by assuming a fixed energy range for the cosmic ray and $\delta=3$, as early noted by Brunetti et al. (1997) and as we did in this work and in Murgia et al. (2009). If, instead, fixed frequency range is used one recovers the classical results that $u_B=(3/4)u_{cr}$ (Pacholczyk 1970).} and the radio emissivity at 1.4 GHz is given by

\begin{equation}
J_{1.4\,GHz}\simeq 8.12\times 10^{-43} (\gamma_{min}/100) B_{\mu G}^{4} ~~~({\rm erg\,s^{-1}cm^{-3}Hz^{-1}}).
\label{Jsynchro}
\end{equation}

In Eq.\,\ref{Jsynchro} we took the isotropy of the electron population into account and we averaged over all the possible
directions between the magnetic field and the line-of-sight, i.e. the field is supposed to be completely tangled on
 an infinitesimally small scale (for a more realistic model see Sect.\,\ref{3D}).
We furthermore assumed that the magnetic field strength scales as a function of the thermal gas density according to

\begin{equation}
B(r)=B_0\left(\frac{n_e(r)}{n_e(0)}\right)^{\eta}
\label{baverage}
\end{equation}
where $B_{0}$ is the average magnetic field strength at the cluster center, while $n_{e}(r)$ is the thermal electron gas density 
profile taken to follow the $\beta$-model (Cavaliere \& Fusco-Femiano 1976):

\begin{equation}
n_e(r)=n_e(0)\left(1+\frac{r^2}{r_c^2}\right)^{-\frac{3\beta}{2}}
\label{ICMmodel}
\end{equation}
where $n_e(0)$ is the thermal gas density at the cluster center and $r_c$ is the cluster core radius.
In the case of A665, the gas density parameters have been taken as derived by Roussel et al. (2000) from ROSAT X-ray data
and rescaled to our chosen cosmology ( $\beta=0.763$, $r_c=112^{\prime \prime}$ and  n$_e$(0)=3.25$\cdot $10$^{-3}$ cm$^{-3}$).

Under these hypotheses, the synchrotron emissivity scales as
\begin{equation}
J_{1.4\,GHz}(r)=J_0\left(1+\frac{r^2}{r_c^2}\right)^{-6\beta \eta}.
\label{Jeta}
\end{equation} 
Since radio halos are optically thin sources, the radio brightness profile results from the integral
 along the line-of-sight of the synchrotron emissivity in Eq.\,\ref{Jeta}, which gives
\begin{equation}
I(r_{\perp})=I_0\left(1+\frac{r_{\perp}^2}{r_c^2}\right)^{-6\beta \eta +0.5}
\label{I}
\end{equation} 
where $I(r_{\perp})$ is the brightness value at the $r_{\perp}$ projected distance from the
cluster center, and $I_0$ is the central radio halo brightness.
By fitting the model in Eq.\ref{I} to the data, we obtain the value of $I_0$ and $\eta$,
 from which $B_0$ can be estimated. We calculated the equipartition magnetic field by neglecting the energy contribution from relativistic protons and other, non-emitting, heavy particles, i.e. we used $\kappa=0$ (where $\kappa$ is conventionally the ratio of the heavy particle to electron energy densities). Adopting $\kappa > 0$ leads to a higher equipartition magnetic field strength: for $\delta=3$, $B \propto (1+ \kappa )^{0.25}$, see also Beck \& Krause (2005). In this respect, the magnetic field values computed in this way should be considered as lower limits.

\begin{figure}[ht]
   \centering
 \includegraphics[width=0.5\textwidth]{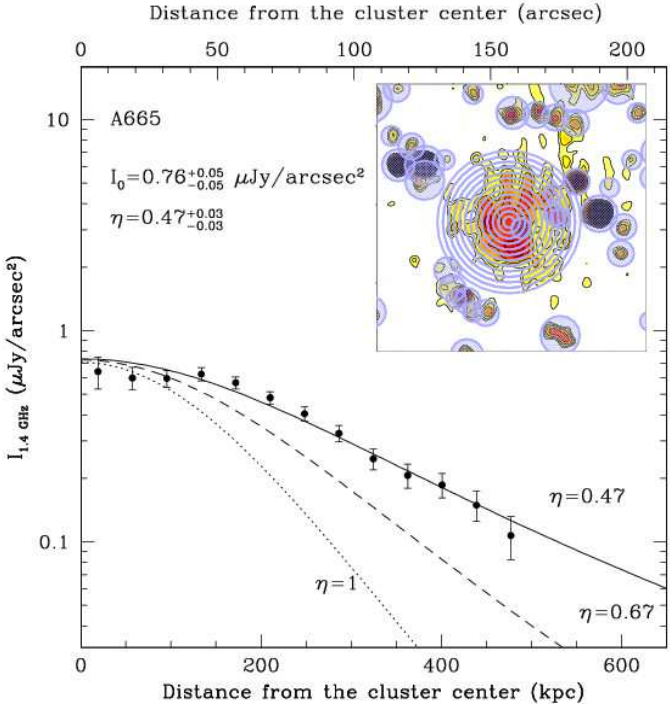}
     \caption{Analytical fit to the observed radio-halo brightness radial profile. Each point represents the azimuthally 
averaged radio brightness at 1.4\,GHz obtained in annuli, as shown in the inset. Discrete sources have been
 excluded from the statistics. The solid line represents the best fit of the magnetic field model described
in the text. The dashed and dot-dashed lines are reference models, see text.}
         \label{profile}
   \end{figure}

In Fig.\ref{profile} we show the azimuthally averaged radio-halo
brightness profile obtained from the C+D configuration image at 25\arcsec resolution. Each data point represents the
average brightness in concentric annuli of half beam width centered on the X-ray peak, as shown in the inset.  
Discrete sources have been masked out and excluded from the statistics. 
The observed brightness profile is traced down to a level 
of 3$\sigma_{I}$ and the best fit of the analytical model in Eq.\,\ref{I} is showed.
The fit was performed in the image plane as described
in Murgia et al. (2009). To properly take the resolution into account, the
model in Eq.\,\ref{I} was first calculated in a
2-dimensional image, with the same pixel size and field of view as that
observed, and then convolved with the same beam by means of a fast
Fourier transform. The resulting image is masked exactly in the same regions 
as for the observations. Finally, the model is azimuthally averaged with the same
set of annuli as were used to obtain the observed radial profile. All these functions were 
performed at each step during the fit procedure.  As a result, the values of the central 
brightness, $I_0$, and the index $\eta$ provided by the fit are deconvolved quantities, and
their estimate includes all the uncertainties related to the masked
regions and to the sampling of the radial profile in annuli of finite
width.

The best fit of the analytical model yields a central brightness
 of $I_0=0.76^{+0.05}_{-0.05}$ $\mu$Jy/arcsec$^2$ and $\eta=0.47^{+0.03}_{-0.03}$, the corresponding equipartition 
magnetic strength at the cluster center is  $B_0=1.58^{+0.04}_{-0.04}$ $\mu$G. Indeed, the value of $\eta$ we
found is very close to the physical situation in which the magnetic field energy density scales as the 
thermal gas density: $B^{2}\propto n_{e}$.

The brightness profile is very sensitive to the value of $\eta$. To show this, we 
fixed the central brightness $I_0$ and we also traced the models corresponding to $\eta=2/3$ (the magnetic field is frozen in 
the thermal gas) and $\eta=1$ (magnetic field radially decreases according to thermal gas density). The two
cases are represented in Fig.\ref{profile}, but these two profiles, however, 
are too steep for the data.

\subsection{Numerical analysis of the magnetic field fluctuations}
\label{3D}
The analytic model presented in the previous section provides a good description of the azimuthally averaged 
brightness radial profile.
However, significant deviations of the diffuse emission from circular symmetry are observed.
These fluctuations can be related to local fluctuations of the intracluster magnetic field strength and direction
around the average. We characterized the power spectrum of these fluctuations following the 
3-dimensional numerical approaches proposed by Tribble (1991) and Murgia et al. (2004).
We simulated a multi-scale random magnetic field in a cubical box. We assumed that the power spectrum 
of the magnetic field fluctuations is a power law with index $n$:

\begin{equation}
|B_k|^2\propto k^{-n}
\label{bpower}
\end{equation}
where the wave number $k$ ranges from $k_{min}$ to $k_{max}$. 
The simulations begin in Fourier space by extracting the amplitude of
the magnetic field potential vector, $\tilde A(k)$, from a Rayleigh distribution whose standard deviation 
varies with the wave number according to $|A_k|^2\propto k^{-n-2}$. 
The phase of the potential vector fluctuations are taken to be completely random. The magnetic field is formed in Fourier space
via the cross product $\tilde B(k)=i k \times \tilde A(k)$. This ensures that the magnetic field is effectively divergence free.
We then performed a 3D fast Fourier transform (FFT) inversion to produce the magnetic field in the real space domain.
The field is then Gaussian and isotropic, in the sense that there is no privileged direction in space for the 
magnetic field fluctuations. The power spectrum normalization is set such that the average magnetic field strength scales as a function of the thermal gas density according to Eq.\,\ref{baverage}. To reduce the computational burden, this operation is 
performed in the real space domain by multiplying the magnetic field by Eq.\,\ref{baverage}. However, to 
preserve the null divergence of the field exactly, this operation should be better performed as a convolution in the Fourier space 
domain before the cross product is formed. Moreover, the convolution with the radial profile in  Eq.\,\ref{baverage} alters
the input power spectrum shape at the edges. Nevertheless, in the specific case 
of A665, the size of the cluster core radius is comparable to the size of the computational grid so that the above effects 
are acceptably negligible.  

This 3-dimensional random magnetic field model is ``illuminated" with the relativistic
electron distribution in Eq.\ref{CRe}, to produce total intensity and polarization synthetic images of the radio halo.
The magnetic field fluctuates in strength and direction from pixel to pixel in the grid.
The integration of the total intensity straightforward: the contributions of the pixels in 
the computational grid are added together by considering the angle between the local magnetic field direction and the line-of-sight. Indeed, the total intensity at a given direction is proportional to the integral along the line-of-sight of the magnetic field projected on the plane of the sky, $B_{\perp}$.

The integration of the polarized intensity is not a simply scalar sum because of two distinct effects. One is that
the magnetic field has random orientations, and the second is that the radio wave polarization plane is subject to the  Faraday rotation
as it traverses the magnetized intracluster medium. Therefore, the integration of the polarized intensity is performed as a vectorial 
sum in which the intrinsic polarization angle of the radiation coming in from the pixels located at a depth $L$ is rotated by an amount 

\begin{equation}
 \Delta\Psi = RM \times \left(\frac{c}{\nu}\right)^2, 
\label{rm}
\end{equation}
where the rotation measure RM is given by
\begin{equation}
RM_{\rm~[rad/m^2]}=812\int_{0}^{L_{[kpc]}}n_{e~[cm^{-3}]}B_{\parallel~[\mu G]}dl.
\label{equaz}
\end{equation}
Here $B_{\parallel}$ is the magnetic field along the line-of-sight.
The combination of the aforementioned effects lead to the so-called internal depolarization of the radio signal.
Finally, to be compared with the data, the synthetic images are convolved with a Gaussian beam as are the observed ones.
The convolution with the beam prevents observing the small-scale fluctuations of the radio halo and causes
 a further suppression of the polarized intensity (beam depolarization).

\begin{figure*}[ht]
   \centering
  \includegraphics[width=15cm]{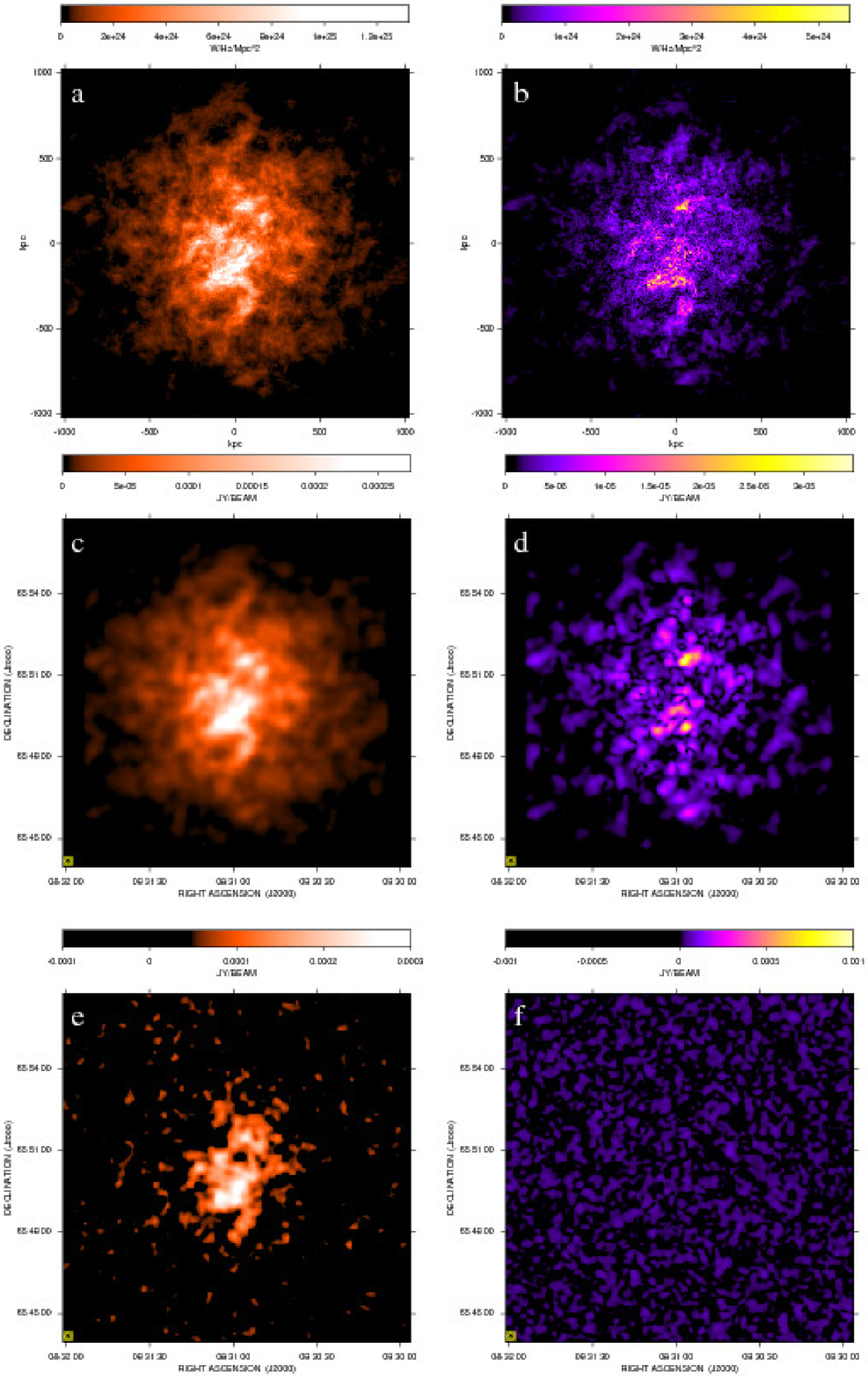}
    \caption{Example of simulated radio halo. Left and right panels refer to the total and polarized intensity, respectively. Top panels: images at full resolution. Mid panels: images at 15\arcsec\, resolution. Bottom panels: as for mid panels but with noise added.}
              \label{sintethyc1}
    \end{figure*}

An example of synthetic A665 radio halo is shown in Fig.\,\ref{sintethyc1}. A turbulent magnetic field
is simulated in a grid of 1024$^3$ pixels with a cell size of 1 kpc/pixel. The magnetic field has a Kolmogorov power spectrum 
slope\footnote{Throughout this paper the power spectra are expressed as vectorial forms in $k$-space.
The one-dimensional forms can be obtained by multiplying by $4\pi k^{2}$ and $2\pi k$ the three and two-dimensional power 
spectra respectively.} $n=11/3$ and fluctuates in the range of spatial scales from $\Lambda_{min}$=4 kpc and $\Lambda_{max}$=512 kpc (where $\Lambda=2\pi/k$). The central magnetic field strength is $\langle B_0\rangle=1.3$ $\mu$G and $\eta=0.47$. The simulated magnetic field is periodic at 
the grid boundaries, so the computational grid has been replicated to reproduce a 
field of view of $2048^2$ kpc$^2$ around the galaxy cluster center. Left and right panels refer to the total and polarized intensity,
 respectively. Top panels refer to the simulations at full resolution and show all the fine structure of the radio halo. The 
expected theoretical fractional polarization for this power spectrum is about 24\%.
Middle panels show the images convolved at 15\arcsec~resolution. The radio halo appears smoother and the fractional polarization is
reduced to about 7\%. Finally, bottom panels show the synthetic radio images with the same noise level as in the observations.  
In total intensity, only the brightest central region of the radio halo are visible, while in polarized intensity the halo 
emission falls below the noise level. This example illustrates how radio halos can be effectively polarized, but because of 
their faintness, detecting this polarized signal is a very hard task with the current radio interferometers.

\section{Simulations results}
\label{simres}
To constrain the power spectrum of the magnetic field fluctuations
in A665, we compared the synthetic radio halo images obtained through the numerical
procedure described in Sect. \,\ref{Characterization of the diffuse radio emission in A665} and
the real radio images and polarization limits presented in Sect.\,\ref{Radio observations and data reduction}.
 Overall, the 3-dimensional magnetic field and relativistic electron models depend on the eight parameters listed in Table \ref{parameters}. For a power-law magnetic field power spectrum with a radial scaling, we have five parameters: $\langle B_0\rangle$, $\eta$, $\Lambda_{min}$, $\Lambda_{max}$, and $n$. As pointed out in Murgia et al. (2004), there are a number of degeneracies between these parameters. In particular, the most relevant to us is the degeneracies between  $\langle B_0\rangle$ and  $\eta$, along with the degeneracy between $\Lambda_{max}$ and $n$. Different combinations of these parameters lead to similar radio halo total intensity and polarization distributions. We decided to concentrate our analysis on $\langle B_0\rangle$ (by fixing $\eta=0.47$ on the basis of the azimuthally averaged radial profile presented in Sect.\,\ref{2D}) and on $\Lambda_{max}$ (by fixing $n=11/3$ on the basis of the Kolmogorov theory for a turbulent medium).
Moreover, we fixed $\Lambda_{min}$=4\,kpc, the minimum allowed by our computational grid. However, we note that a higher $\Lambda_{min}$ has a negligible impact on the simulation results since, for a Kolmogorov spectral index, most of the magnetic field power is on larger scales. 
The random magnetic field is illuminated with a population of relativistic electrons 
with Lorentz factors between $\gamma_{min}=100$ and $\gamma_{max}=+\infty$. The electrons spectrum is a power law 
with index $\delta=3$, according to the observed radio-halo spectral index  $\alpha_{0.3}^{1.4}\simeq 1$ (see Sect.\,\ref{Abell665}), 
and its energy density is in local equipartition with the magnetic field energy density.

Because of the random nature of the intracluster magnetic field, the comparison between synthetic and 
observed images requires several simulation runs with different seeds for each given couple of the fitting parameters 
$\langle B_0\rangle$ and $\Lambda_{max}$, this because the same set of the magnetic field power spectrum parameters 
results in a different realization of the same synthetic halo. Indeed, we chose to simulate 10 equally spaced  values 
of the minimum wave number from $k_{min}=0.0061$\,kpc$^{-1}$ to $k_{min}=0.061$\,kpc$^{-1}$.
Correspondingly, the outer scale of the magnetic field fluctuations ranges from $\Lambda_{max}=1024$ kpc 
(the maximum allowed by our computational grid) down to $\Lambda_{max}=102.4$ kpc (i.e. very close to the linear 
resolution of the radio images). For each value of $\Lambda_{max}$, we realized 10 more simulation runs, with different seeds,
 in which we determined the value of $\langle B_0\rangle$ by fitting the synthetic radio halo brightness profile to the 
observed one. Indeed, we realized a total of 100 simulations by running the FARADAY code (Murgia et al. 2004) in the 
Cybersar-OAC computer cluster.

\begin{table*}
\caption{Parameters adopted in the radio halo simulation.}             
\label{parameters}      
\centering          
\begin{tabular}{c c c }    
\hline\hline       
    Parameter     & Value                    & Description          \\ 
\hline                        
  $\langle B_0\rangle$           &{\it free}        & magnetic field strength at the cluster center\\
  $\eta$          &0.47                       & magnetic field radial decrease, $\langle B(r)\rangle=\langle B_0\rangle\left(\frac{n_e(r)}{n_e(0)}\right)^{\eta}$\\
  $n$             &11/3                      & magnetic field power spectrum index, $|B_k|^2 \propto k^{-n}$\\
  $\Lambda_{min}$ &4\,kpc                      & minimum scale of fluctuation, $\Lambda_{min}=2\pi/k_{max}$\\
  $\Lambda_{max}$ &{\it free}                       & maximum scale of fluctuation, $\Lambda_{max}=2\pi/k_{min}$\\
  $\gamma_{min}$  &100                       & minimum relativistic electron Lorentz factor\\
  $\gamma_{max}$  &$+\infty$                  & maximum relativistic electron Lorentz factor\\
  $\delta$       &3                          & power-law index of the energy spectrum of the relativistic electrons \\

\hline                                 
\end{tabular}
\end{table*}

\begin{figure*}[ht]
   \centering
  \includegraphics[width=15cm]{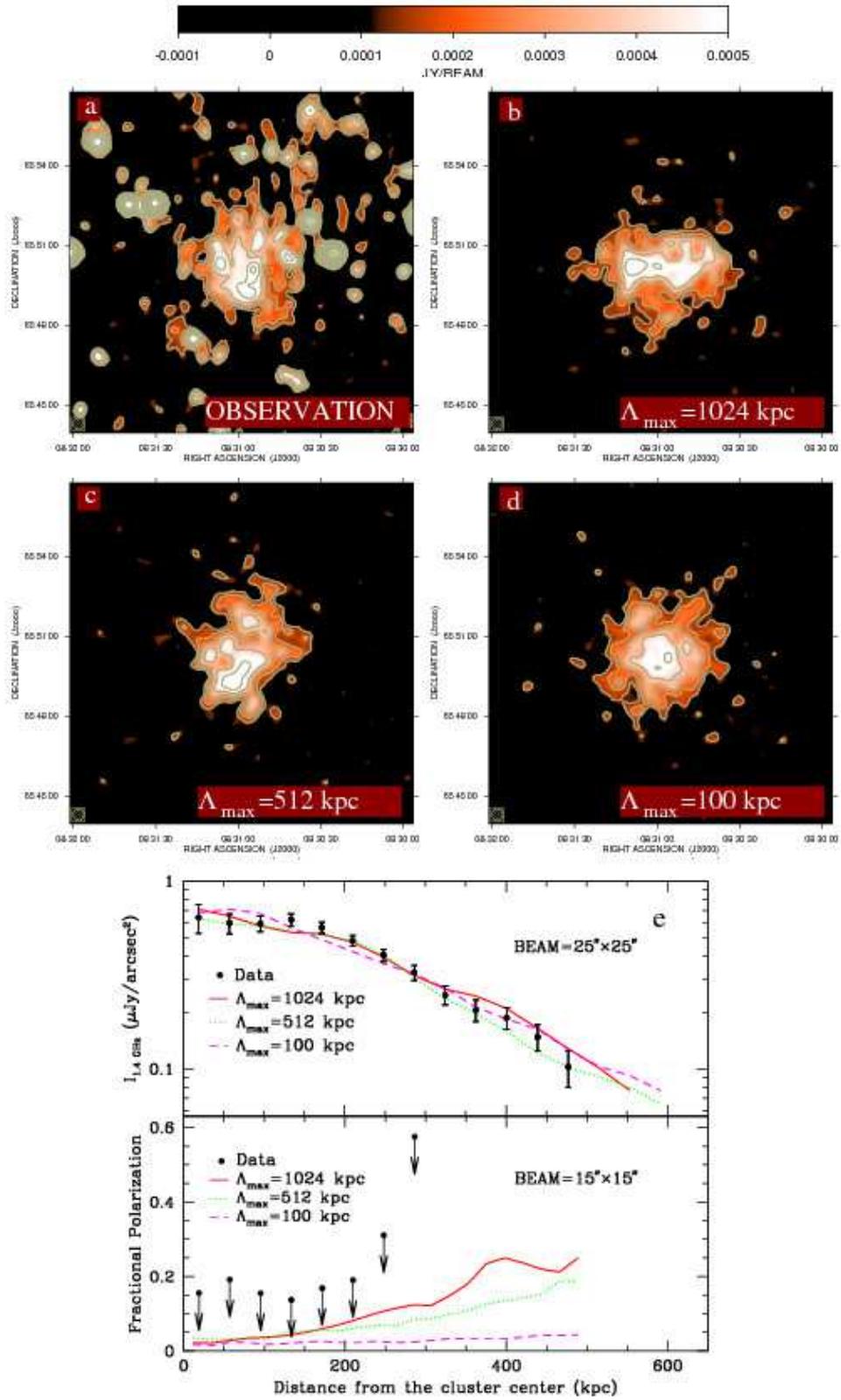}
   \caption{ Observed (a) and simulated (b, c, d) surface brightness images, with FWHM of 25$\arcsec \times$ 25$\arcsec$. Both the observation and simulation first contour levels are drawn at 3$\sigma$ and the rest are spaced by a factor $\sqrt{2}$. The simulations correspond to three different $\Lambda_{max}$: 1024, 512, 
and 100 kpc.  In the bottom panels (e), the radial profiles of the radio brightness and the fractional polarization are shown. The polarization upper limits are at 1$\sigma$ level.}
              \label{simulations}
    \end{figure*}

In Fig.\,\ref{simulations} we show the observed radio halo in total emission at 25\arcsec\,resolution (panel a) along with 
 the synthetic radio-halo images corresponding to three different values $\Lambda_{max}$ (panels b, c, and d).
The synthetic images have the same resolution and noise level as the observed one and were obtained with the 
procedure described in Sect.\,\ref{3D}. First glance, the synthetic images look quite similar to 
the observed one in terms of total extension and average brightness.
In fact, the top panel of Fig.\,\ref{simulations}e shows that all three simulations
 fit the total intensity radial profile remarkably well at 1.4\,GHz of the observed A665 radio-halo image at 
25\arcsec\,resolution. The different lines in the bottom panel of Fig.\,\ref{simulations}e represent the simulated fractional
 polarization at different radii from the cluster center as expected at 15\arcsec resolution before the noise is applied.
The model fractional polarization is consistent with the upper limits derived on the base of the VLA C configuration
 observation presented in this work. The best-fit central magnetic field strength for the three simulations is
  around $\langle B_0\rangle$ =1.3 $\mu$G, in good agreement with the value found on the basis of the 2-dimensional 
 analytical fit in Sect.\,\ref{2D}.

Although the examples of the synthetic halos shown in Fig.\,\ref{simulations} visually appear similar to the
observed halo, it is clear that increases as $\Lambda_{max}$, and the magnetic field power approaches as larger scales, 
the simulated radio halos change shape. In particular, when the outer scale of the magnetic field fluctuations
 is close to the observing beam, the halo is smooth and rounded (Fig.\,\ref{simulations}d). Increasing $\Lambda_{max}$
 results in a much distorted radio halo morphology and in a significant offset of the radio halo peak from the 
cluster centered (particularly evident in Fig.\,\ref{simulations}b). The  simulated 
 clusters are centered on the observed X-ray peak. 
 
We indeed tried to evaluate quantitatively the value of $\Lambda_{max}$, which best reproduce the observed
 intensity fluctuations of the radio halo, as originally proposed time ago by Tribble (1991). To do this, 
we analyzed the residual images obtained by fitting and then subtracting, the analytical model in Eq.\,\ref{I} to both the simulated and the observed halo images. Since the analytical model has a perfect circular symmetry,
 it is particular well-suited to highlighting the halo fluctuations around the average profile.

\begin{figure*}[ht]
   \centering
 \includegraphics[width=17cm]{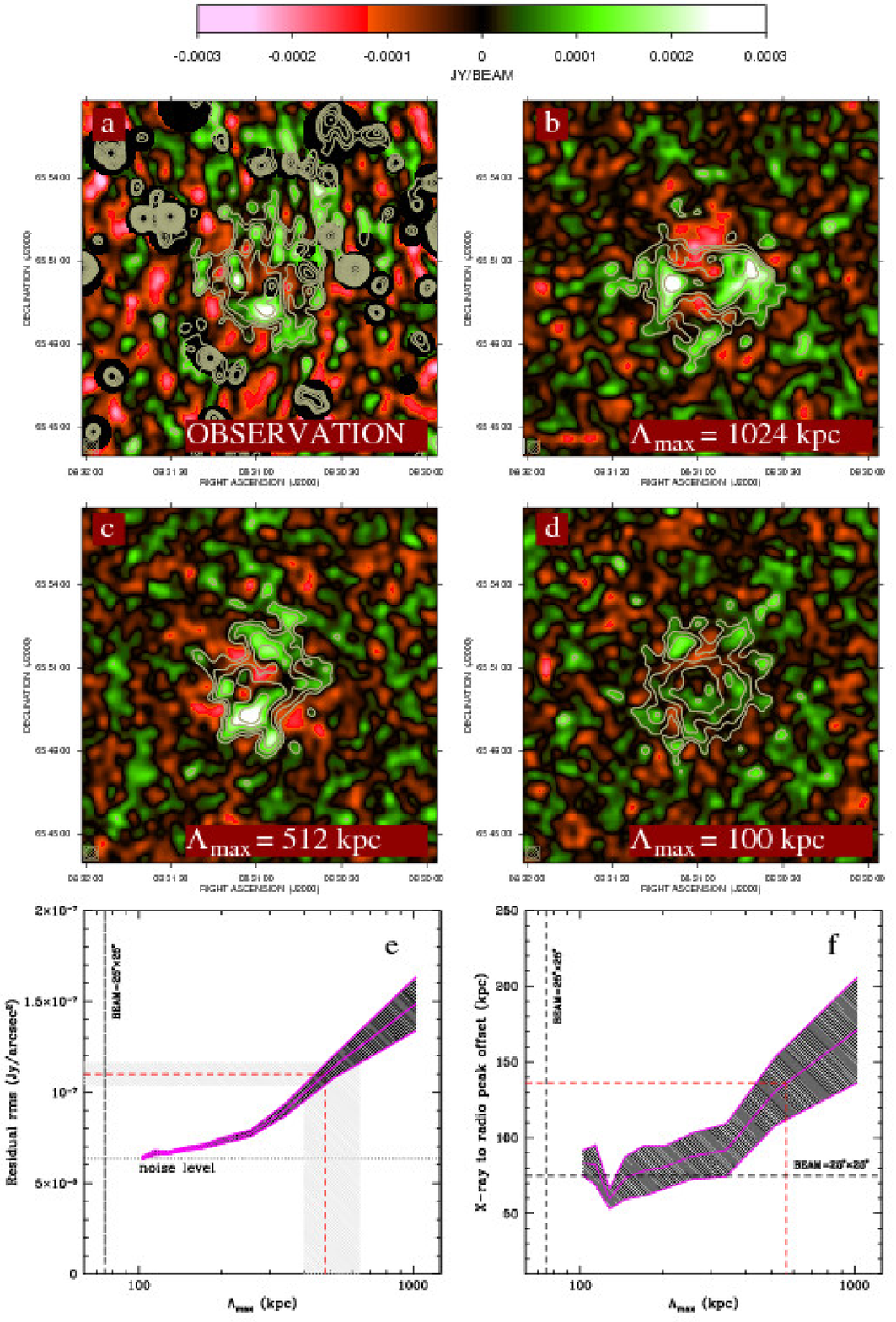}
    \caption{ Observed (a) and simulated (b, c, d) residuals images, with FWHM of 25$\arcsec \times $25$\arcsec$. Both the observation and simulation first contour levels are drawn at 3$\sigma$ and the rest are spaced by a factor $\sqrt{2}$. Panel e) and f) show the residuals rms 
versus $\Lambda_{max}$ and the X-ray to radio peak offset versus $\Lambda_{max}$, respectively.}
    \label{residuals}
    \end{figure*}

In the top panels of Fig.\ref{residuals} we show residuals images corresponding to the images 
shown in Fig.\,\ref{simulations} with the total intensity iso-contours overlapped. The higher  $\Lambda_{max}$, the 
 higher the residual levels and the larger the fluctuation patches.

In Fig.\,\ref{residuals}e the root-mean-square of the residual images, evaluated inside the inner $\sim 200$ kpc from the cluster center, is plotted against $\Lambda_{max}$. The solid line and the shaded region represent the mean and 
the standard deviation of the mean derived from the simulations, respectively. First of all, the 
residuals approach the noise level as  $\Lambda_{max}$ approaches the observing beam. This result implies that, 
if $\Lambda_{max}$ is smaller than $\lesssim 100$ kpc, the expected rms of the halo intensity fluctuations is already below the
 noise level of the observations $\sigma_{I}=0.06$\,$\mu$Jy/arcsec$^2$. As the maximum scale of the intracluster magnetic field
 power spectrum increases, the rms level
 of the radio halo fluctuations increases, reaching a value as high as 0.15\,$\mu$Jy/arcsec$^2$ for $\Lambda_{max}=1024$\,kpc.

By comparing the observed fluctuation rms level of 0.11\,$\mu$Jy/arcsec$^2$ with the 
simulated trend we estimate that the outer scale of the magnetic field fluctuations is  
$\Lambda_{max}\simeq 400\div 500$ kpc (i.e. $k_{min}=0.014 \div 0.012$ kpc$^{-1}$). 
In Fig.\,\ref{residuals}f the offset between the X-ray and radio peak is plotted against $\Lambda_{max}$. This 
indicator has a larger dispersion but is much more direct since it does not involve a specific handling of the observed and 
simulated images. As observed in Fig.\,\ref{simulations}, the offset of the radio peak from the cluster center (here
assumed to be coincident with the X-ray peak) increases with  $\Lambda_{max}$. By comparing the observed 
offset of 135 kpc with the simulated trend, it turns out that  $\Lambda_{max}$ should be about $500\div 600$\,kpc in agreement 
with the residual analysis.

To summarize, the results of the 3-dimensional simulations indicate that the radio halo emission in A665 is 
consistent with a central magnetic field strength of about 1.3 $\mu$G. To explain the azimuthally averaged radio brightness profile, the magnetic field energy density should decrease following the thermal gas density, leading to an averaged magnetic field strength over the central 1 Mpc$^3$ 
of about 0.75 $\mu$G. From the observed brightness fluctuations of the radio halo, we infer that the outer scale
of the magnetic field power spectrum is $\sim$450 kpc, the corresponding magnetic field auto-correlation 
length\footnote{Following En{\ss}lin \& Vogt (2003), we define the magnetic field auto-correlation length as 
\begin{equation}
\Lambda_{B}=\pi\cdot\frac{\int_{0}^{+\infty} |B_{k}|^{2}k\,dk}{\int_{0}^{+\infty} |B_{k}|^{2}k^2\,dk}.
\end{equation}} 
is $\Lambda_{B}\simeq$100 kpc.
 
\section{The depolarization of discrete radio sources}
\label{Radio galaxy emission in A665}
   \begin{figure*}[ht]
   \centering
  \includegraphics[width=18cm]{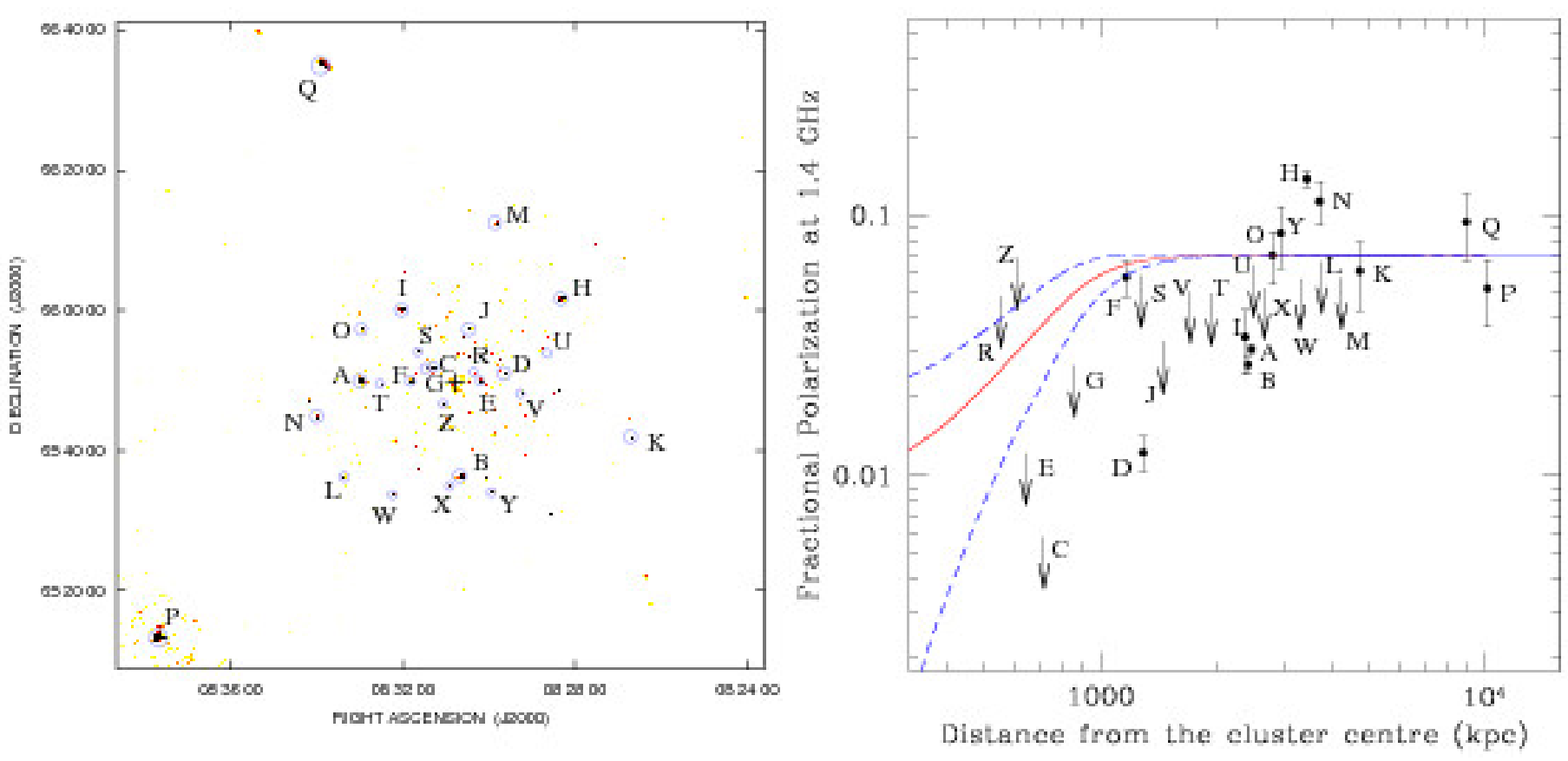}
      \caption{Depolarization of the brightest discrete radio sources as a function of the distance from the cluster center (marked by the cross). 
The upper limits are at the $3\sigma$ level. The solid and dashed lines respectively represent the mean and the dispersion of the depolarization trend 
expected on the basis of the magnetic field model that best reproduces the radio halo in A665, see text for details.}
              \label{Radiogalassie}
    \end{figure*}

In this section we investigate the effects of the intracluster magnetic field in A665 on the polarization
 properties of the discrete radio sources belonging to the cluster of galaxy itself or situated in background to it (see Table\,\ref{tabella_radiog}).
In particular, it is expected that the radio sources whose emission crosses the central region of the cluster
 suffer a higher Faraday rotation, hence a higher beam depolarization, with respect to the sources at 
larger impact parameters (see e.g. Murgia et al. 2004). In the left panel of Fig.\,\ref{Radiogalassie}, we show a field of view of about $1.5^{\rm o}\times 1.5^{\rm o}$ around A665 taken 
from a 15\arcsec\,resolution image. All the discrete sources brighter than 500 $\mu$Jy/beam (20$\sigma_{I}$) are labeled. 
This severe cut guarantees a maximum error on the fraction polarization of 0.03.

For each of these sources we plotted the fractional polarization against their distance from the cluster center (right panel of Fig.\ref{Radiogalassie}). At large impact parameters, $r_{\perp}>2000$ kpc, the observed 
radio sources fractional polarization at 1.4 GHz and 15\arcsec\,resolution oscillates around 0.1. The plot also
 shows a hint that the discrete source seen in projection close to the cluster center is more 
depolarized.

The continuous line is the expected depolarization trend calculated on the basis of the best magnetic field
 model derived from the fit of the radio halo image (see Sect.\ref{simres}). The dashed lines represent the
dispersion of the model. The depolarization model assumes that all the considered sources are situated at
the cluster mid-plane and that the intrinsic polarization is fully ordered. As intrinsic value we assumed the mean of the fractional polarization of the radio galaxies located at more than 2\,Mpc from the cluster center.

Although the relatively large scatter of the data and the simplicity of the model prevent us from deriving other useful constraints to the magnetic field properties
in A665, we note that the model prediction broadly agrees with the observed depolarization trend. Studies at possibly higher resolution and sensitivity are required to measure the fractional polarization 
for many more sources so that this method can be applied successfully.

\begin{table*}[ht]
\caption{A665 radio galaxies used to study signal depolarization
  through intracluster medium.}             
\label{tabella_radiog}      
\centering          
\begin{tabular}{c c c c c c }     
\hline\hline       
Source &        RA     &      DEC    &   $r_\perp$   &  $FPOL$     &       $S_{1.4\,GHz}$       \\   
~      &     (J2000)   &    (J2000)  & (kpc) &  (\%)       &    ($mJy$)    \\
\hline
A  &  08:33:12 & 65:50:41   &2474 &      3.0 $\pm$ 0.1   &    222$\pm$ 7  \\ 
B  &  08:30:53 & 65:36:60   &2429 &      2.7 $\pm$ 0.2   &    122$\pm$ 4\\               
C  &  08:31:31 & 65:52:36   &712  &          $<0.6$      &    35$\pm$ 1\\               
D  &  08:29:51 & 65:51:53   &1293 &      1.2 $\pm$ 0.2   &    41$\pm$ 1 \\             
E  &  08:30:25 & 65:50:42   &639  &          $<1.0$      &    12.8$\pm$  0.4\\              
F  &  08:32:02 & 65:50:38   &1166 &      5.8 $\pm$ 0.9   &    6.4$\pm$ 0.2\\              
G  &  08:31:40 & 65:52:33   &855  &          $<3$        &    3.8$\pm$ 0.1\\              
H  &  08:28:34 & 66:02:17   &3449 &     13.8 $\pm$ 1.0   &    27.2$\pm$ 0.8\\             
I  &  08:32:17 & 66:00:50   &2382 &      3.4 $\pm$ 0.9   &    12.0$\pm$ 0.4\\              
J  &  08:30:42 & 65:58:09   &1456 &          $<3$        &    2.8$\pm$ 0.09\\             
K  &  08:26:57 & 65:42:31   &4745 &      6.1 $\pm$ 1.8   &    19.9$\pm$ 0.6\\             
L  &  08:33:34 & 65:37:01   &3779 &       4.0$\pm$2.2    &    5.9$\pm$ 0.2\\             
M  &  08:30:04 & 66:13:06   &4259 &      3.2 $\pm$ 1.9   &    5.7$\pm$ 0.2\\             
N  &  08:34:13 & 65:45:25   &3728 &     11.3 $\pm$ 2.2   &    3.7$\pm$ 0.1\\            
O  &  08:33:11 & 65:58:02   &2810 &      7.0 $\pm$ 1.6   &    2.8$\pm$ 0.09\\  
P  &  08:37:44 & 65:13:31   &10203&       5.2$\pm$1.5    &    1275$\pm$2\\
Q  &  08:34:14 & 66:35:52   &8991 &       9.4$\pm$2.7    &    341$\pm$2\\
R  &  08:30:34 & 65:51:54   &550  &          $<5$        &    1.05$\pm$0.04\\
S  &  08:31:52 & 65:54:53   &1277 &          $<6$        &    0.77$\pm$0.03\\
T  &  08:32:44 & 65:50:15   &1946 &          $<5$        &    1.77$\pm$0.06\\
U  &  08:28:51 & 65:54:36   &2512 &          $<6$        &    0.81$\pm$0.03\\
V  &  08:29:28 & 65:49:08   &1713 &          $<5$        &    1.19$\pm$0.04\\
W  &  08:32:26 & 65:34:28   &3313 &          $<6$        &    3.2$\pm$0.1\\
X  &  08:31:08 & 65:35:39   &2676 &          $<5$        &    2.31$\pm$0.07\\
Y  &  08:30:09 & 65:34:57   &2952 &       8.5$\pm$2.2    &    1.22$\pm$0.04\\
Z  &  08:31:16 & 65:47:30   &607  &          $<7$        &    0.42$\pm$0.02\\         
 \hline 
 \multicolumn{6}{l}{\scriptsize Col. 1: Radio galaxy labels; Col. 2 to Col. 3: Source's position;}\\
\multicolumn{6}{l}{\scriptsize Col. 4: Distance from cluster center; Col. 5: Polarization Percentage ;}\\    
\multicolumn{6}{l}{\scriptsize Col. 6: Flux density.}\\      
\end{tabular}
\end{table*}

\section{Conclusions}
In this work we presented a new deep VLA observation at 1.4 GHz of the cluster of galaxies A665.
By combining this observation with a previous VLA observation at lower resolution, we studied the intracluster magnetic-field power spectrum 
by analyzing the radio halo brightness fluctuations, following both an original idea by Tribble (1991) and the numerical 
approach proposed by Murgia et al. (2004). 

Our findings are summarized as follows. We simulated Gaussian random three-dimensional turbulent magnetic-field models in order to reproduce the observed radio halo emission. By comparing observed and synthetic radio 
halo images, we constrained the strength and structure of the intracluster magnetic field. We assumed that the magnetic 
field power spectrum is a power law with a Kolmogorov index, and we imposed a local equipartition of energy density 
between relativistic particles and the magnetic field. Under these assumptions, we find that the radio halo emission in A665 is 
consistent with a central magnetic field strength of about 1.3 $\mu$G. To explain the azimuthally averaged radio brightness profile, the magnetic field energy density should decrease following the thermal gas density,
leading to an averaged magnetic field strength over the central 1 Mpc$^3$ of about 0.75 $\mu$G. From the observed 
brightness fluctuations of the radio halo, we inferred that the outer scale of the magnetic field power spectrum is $\sim$ 450 kpc, the corresponding magnetic field auto-correlation length is 100 kpc. We also
find a hint that the discrete source seen in projection close to the cluster center is more 
depolarized. The best-fit magnetic model broadly agrees with the observed depolarization trend.

More tightening constraints could be potentially obtained by detecting the radio halo polarization fluctuations, not just
 total intensity fluctuations. In fact, the ratio of two former quantities, i.e. the fractional polarization, is
a very robust indicator of the intracluster magnetic-field power spectrum, because it only marginally depends on
 the shape of the energy spectrum of the synchrotron electrons and on the equipartition assumption.
Therefore, it would be very important to improve the sensitivity of the future observations in order to detect 
polarized signal in the most radio halo possible. This is a science case in the new generation of instruments 
in radio astronomy.

\begin{acknowledgements}
We thank the referee for very helpful comments that led to improvements in this work.
This work is part of the ``Cybersar'' Project, which is managed by the COSMOLAB Regional Consortium with
the financial support of the Italian Ministry of University and Research (MUR), in the context of the ``Piano Operativo Nazionale
  Ricerca Scientifica, Sviluppo Tecnologico, Alta Formazione (PON  2000-2006)''. The research was partially supported by ASI-INAF I/088/06/0 -
High Energy Astrophysics and PRIN-INAF2008. The National Radio Astronomy Observatory (NRAO) is a facility of the National Science Foundation,
operated under cooperative agreement by Associated Universities, Inc. We are grateful to Antonietta Fara and
 Riccardo Pittau for their assistance with the Cybersar-OAC computer cluster. 

\end{acknowledgements}

\end{document}